# An investigation of the temperature dependency of the relative population inversion and the gain in EDFAs by the modified rate equations

**Cüneyt Berkdemir and Sedat Özsoy**

*Department of Physics, Faculty of Science and Literature, Erciyes University, 38039 Kayseri, Turkey*

*Corresponding author: e-mail: berkdemir@erciyes.edu.tr*

**Abstract**

The dependence of the relative population inversion in $Er^{3+}$-doped fiber amplifiers (EDFAs) upon temperature and cross−sections, taking into account the amplified spontaneous emission (ASE), are investigated theoretically by the modified rate equation model for 980 *nm* and 1470 *nm* pumping conditions. For the temperature range from 0 to +50 $^{o}C$ and at the different signal wavelengths, the temperature and cross section−dependent gain characteristics with respect to pump powers are also examined in detail for the both conditions. As a consequence, the dependence of the performance of EDFAs on temperature for 980 *nm* pumping is weaker than that for 1470 *nm* pumping, not only at room temperature but also at the temperature range of 0 to +50 $^{o}C$. However, the performance of EDFAs is more efficient at the pumping wavelength of 1470 *nm* than that of 980 *nm* for a wide range of temperature and high-pump powers. The results of this theoretical model are a good agreement with the experimental ones in the literature.

## 1. Introduction

The dependence of the gain in $Er^{3+}$-doped fiber amplifiers (EDFAs) upon temperature is important for the system applications such as post amplifiers, in-line amplifiers and preamplifiers. In addition, emission and absorption cross-sections also affect the gain performance of EDFAs. Fuchtbauer-Ladenberg (FL) equation commonly depends on the experimental values obtained from fluorescence and absorption bandwidths. Whereas, McCumber theory, which is developed for the study of phonon-terminated lasers based upon transition metal ions [1], gives a more accurate expression for the reciprocity relation between absorption and emission cross sections than FL equation. McCumber relation depends on temperature and requires an experimental value of emission cross section.

The theoretical investigation of EDFAs by using the rate equations is based on the determination of the population difference between amplification levels. The population difference depends on the parameters such as normalized pump photon flux (pump power), normalized signal photon flux (signal power), Boltzmann factor and cross-sections ASE (or noise figure representation), as well as $Er^{3+}$ concentration. Mao et al. and Jaraba et al. obtained analytical solutions by including cross-section factors [2-3] but they are for the room temperature. In recent works, using these solutions, the influence of erbium-doped fiber length on the gain characteristics of the EDFA has been analyzed [4-5].

In this work, the rate equations are modified by taking temperature and cross section terms into account to investigate the gain characteristics of EDFAs for a wide range of temperature. In addition, the contribution of ASE is also included in the study. A 1 % change in the gain coefficient of the EDF with the temperature corresponds to a 0.3 dB gain difference in a typical EDFA with 30 dB of erbium gain. This change is a significant part of the available gain in an EDFA. Hence, the temperature dependency of the erbium ions is an important part of the overall temperature dependence of an EDFA. From the point of view, the modified rate equation model for predicting the temperature dependence of a particular EDFA will be given in the next section.

## 2. Descriptions and The theoretical model

The dynamic behavior of EDFAs can be explained by a three-level system which consists of ground, metastable and upper levels. The amplification process depends basically on the relative population inversion of the doping ions and the fluorescence lifetime of their exited state in the amplification medium. When the fiber is doped with $Er^{3+}$ ions or with ions of other rare-earth elements, the fiber can be operated as an active amplification medium that has to be pumped by the absorption of light from the pump source at either 980 *nm* or 1470 *nm*.

Fig. 1 shows the energy diagram of erbium ions in glass hosts such as silica, telluride and fluorozirconate.

**Fig.1.**

The rate equations for the system are,

$$\frac{\partial}{\partial t} N_3 = R_p (N_1 - N_3) - \gamma_{32} N_3, \tag{1}$$

$$\frac{\partial}{\partial t} N_2 = - S_{21} N_2 + S_{12} N_1 - \gamma_2 N_2 + \gamma_{32} N_3, \tag{2}$$

$$\frac{\partial}{\partial t} N_1 = - R_p (N_1 - N_3) + \gamma_2 N_2 - S_{12} N_1 + S_{21} N_2. \tag{3}$$

Where, $S_{12,21}$ denote the stimulated absorption and emission rates associated with the signal between the first and the second (metastable) level $R_p$ refers to both the pumping rate from the first to the third level and the stimulated emission rate between them. $N_1$, $N_2$ and $N_3$ are population of $Er^{3+}$-ions in first, second and third level, respectively. $\gamma_2$ is the transition probability, which is the sum of the spontaneous radiative and nonradiative transition probabilities from the metastable to the ground level. There are two possibilities of decay from the exited state (level 3) to the lover states: the spontaneous radiative and nonradiative transition. In practice, for the most typical cases, it is assumed that the spontaneous nonradiative transition (from $^4I_{11/2}$ to $^4I_{13/2}$) dominates on the radiative transitions (from $^4I_{11/2}$ to $^4I_{13/2}$ and $^4I_{11/2}$ to $^4I_{15/2}$). Thus, the transition probability per unit time ($\gamma_{32}$) for an $Er^{3+}$ ion represents the thermal relaxation from the upper state (third level) to the lower level (second level). For the transition $^4I_{13/2} \to ^4I_{15/2}$, it is assumed that the spontaneous radiative transition is more probable than the spontaneous nonradiative transition, because the lifetime of the latter is very short compared with the lifetime ($\gamma_2^{-1} = \tau = 10$ *msec*) of the Er$^{3+}$ ion in the metastable level. Here, $N_1 + N_2 + N_3 = N$ (constant) is the total number of Er$^{3+}$ ions in the system with the three-level. For 1470 *nm* and 980 *nm* pumping configurations, the rate equations are solved under the conditions of steady state regime, where all of the level populations are time invariant, i.e., $\partial N_i / \partial t = 0$ ($i = 1, 2, 3$).

The relation between the mean numbers of atoms in the two levels placed closely of a system being at thermal equilibrium are given by Boltzmann's distribution [6] as

$$\beta = \frac{N_3}{N_2} = \frac{g_3}{g_2} \exp(-\Delta E_3 / k_B T). \tag{4}$$

Where $\Delta E_3 = E_3 - E_2$, the energy difference between the third and the second level, $\beta$ is the Boltzmann population factor and $k_B$ is the Boltzmann's constant ($k_B = 1,38 \times 10^{-23}$ *J/K*). The degeneracy of the ground level of Er$^{3+}$ ion, $g_1$, has 8 ($^4I_{15/2}$) components (eightfold); $g_2$ (for the metastable level) has 7 ($^4I_{13/2}$) components (sevenfold) and $g_3$ (for the excited level) has 6 ($^4I_{11/2}$) components (sixfold).

The relative population difference ($\Delta N / N = (N_2 - N_1)/N$) for 1470 *nm* pumping configuration can be obtained from Eqs.(1-3) as

$$\frac{\Delta N}{N} = \frac{R_p \tau (1 - \beta) + S_{12} \tau (1 - \eta) - 1}{R_p \tau (1 + 2\beta) + S_{12} \tau (1 + \beta + \eta) + 1}, \tag{5}$$

with,

$$N_1 = \frac{\beta R_p \tau + S_{12} \tau \eta + 1}{R_p \tau (1 + 2\beta) + S_{12} \tau (1 + \beta + \eta) + 1} N, \tag{6}$$

$$N_2 = \frac{R_p \tau + S_{12} \tau}{R_p \tau (1 + 2\beta) + S_{12} \tau (1 + \beta + \eta) + 1} N, \tag{7}$$

and

$$\eta = \frac{S_{21}}{S_{12}}. \tag{8}$$

Where it is assumed that $N_3 = \beta N_2$, because of $N_2$ and $N_3$ levels of Er$^{3+}$ ion pumped at 1470 *nm* wavelength are spaced closely, and the fast relaxation process. Thus, the $\beta$ parameter is inserted into the rate equations for the case of 1470 *nm* pumping. In addition, there is a relationship between the transition rates $S_{12}$ and $S_{21}$ as fallows [7],

$$S_{12} = \frac{\Gamma_s \sigma_{12}}{h \nu_s A} (P_s + P_{ASE}^+ + P_{ASE}^-), \tag{9}$$

$$S_{21} = \frac{\Gamma_s \sigma_{21}}{h\nu_s A}(P_s + P_{ASE}^+ + P_{ASE}^-), \tag{10}$$

$$S_{21} = \frac{\sigma_{21}(\lambda_S)}{\sigma_{12}(\lambda_S)} S_{12} = \eta(\lambda_S) S_{12}. \tag{11}$$

Where $\Gamma_S$ is the mode overlap factor, $A$ is the effective cross-sectional area of the fiber core, $P_s$ is the signal power coupled into the fundamental fiber mode $LP_{01}$ and $P_{ASE}^{\pm}$ is the amplified spontaneous emission power, which are the superscript "+" designates ASE copropagating with the signal, and "–" when it counterpropagate to the signal. The dependence on longitudinal coordinate $z$ of the pump, signal and ASE powers are given as follow

$$\frac{dP_s^{\pm}}{dz} = \pm \Gamma_s (\sigma_{21} N_2 - \sigma_{12} N_1) P_s^{\pm}, \tag{12}$$

$$\frac{dP_p^{\pm}}{dz} = \pm \Gamma_p (\sigma_{31} N_2 - \sigma_{13} N_1) P_p^{\pm}, \tag{13}$$

$$\frac{dP_{ASE}^{\pm}}{dz} = \pm 2h\nu_s \Delta\nu \sigma_{21} \Gamma_s N_2 \pm \Gamma_s (\sigma_{21} N_2 - \sigma_{12} N_1) P_{ASE}^{\pm}, \tag{14}$$

where $\sigma_{13}$ and $\sigma_{31}$ are the pump absorption and the pump emission cross sections, respectively, $\sigma_{12}$ and $\sigma_{21}$ are the signal absorption and the signal emission cross sections, respectively, and $\Delta\nu$ is the amplifier homogeneous bandwidth. At the signal wavelength $\lambda_S$, $\sigma_{12}(\lambda_S)$ are connected with $\sigma_{21}(\lambda_S)$ via McCumber's relation [8],

$$\sigma_{21}(\lambda_S) = \sigma_{12}(\lambda_S) e^{(\varepsilon - h\nu)/k_B T}, \tag{15}$$

where $h$ is the Planck's constant and $\varepsilon$ is the temperature-dependent excitation energy which is the net free energy required for exciting one Er$^{3+}$ ion from the $^4I_{15/2}$ to $^4I_{13/2}$ level at the temperature $T$. It can be rewritten in the following form that its accuracy is very high (error < 3 %, [9]):

$$\frac{\sigma_{21}(\lambda_S)}{\sigma_{12}(\lambda_S)} = exp\left(\frac{hc}{k_B T}\left[\frac{1}{\lambda_0} - \frac{1}{\lambda_S}\right]\right), \tag{16}$$

where $\lambda_0$ is a constant that depends on the details of electronic structures belonging to the both ground and the exited state of the erbium-ion and to be calculated. The relative population difference for 980 *nm* pumping configuration is also obtained from the rate equations as

$$\frac{\Delta N}{N} = \frac{R_p \tau + S_{12} \tau (1-\eta) - 1}{R_p \tau + S_{12} \tau (1+\eta) + 1}, \tag{17}$$

with,

$$N_1 = \frac{S_{12}\tau\eta + 1}{R_p\tau + S_{12}\tau(1+\eta) + 1} N, \tag{18}$$

$$N_2 = \frac{R_p\tau + S_{12}\tau}{R_p\tau + S_{12}\tau(1+\eta) + 1} N. \tag{19}$$

When $\gamma_{32}$ is large compared to the pumping rate into level 3, the population $N_3$ is very close to zero, so the total population equals $N_1$ plus $N_2$.

$\Delta N$, that is given by Eq.(5) for 1470 *nm* pumping and by Eq.(17) for 980 *nm* pumping, is a very important factor, because the gain of EDFAs depends firmly on the relative population inversions. If the excitations for the both pumping configurations are very strong, then it can be written from Eq.(5) and Eq.(17).

$$R_{p(1470)} > \frac{1 - S_{12}\tau(1 - \eta_{(1470)})}{(1-\beta)\tau}, \tag{20a}$$

$$R_{p(980)} > 1 - S_{12}\tau(1 - \eta_{(980)}). \tag{20b}$$

When the pumping rates or the pumping powers required for the excitation are extremely strong, the population differences in Eq.(5) and Eq.(17) become,

$$\lim_{R_{p(1470)} \to \infty} \Delta N = \left(\frac{1-\beta}{1+2\beta}\right) N, \tag{21}$$

$$\lim_{R_{p(980)} \to \infty} \Delta N = N. \tag{22}$$

These equations verify that, giving the same results $N_3$ is negligible when the value of $\beta$ in Eq.(4) is very small. Under these conditions, the greater population inversion and hence the stronger amplification effect are obtained as expected.

For the strong pumping, the various transitions to the upper levels can be seen, which is so called the excited stated absorption (ESA). There is no ESA for 980 and 1470 *nm* pump wavelengths whereas it occurs for 510, 532, 665, 810 *nm* pumping wavelengths and reduces the efficiency of EDFAs. Thus, EDFAs pumped at the wavelengths used here show not only the higher gain per unit pump power, but also the higher signal output power and the lower amount (a few decibels) of amplified spontaneous emission (ASE) noise [10]. The net signal gain *G(λ)* is defined as fallows [11]:

$$G(\lambda) = e^{\gamma \cdot L}, \tag{23}$$

$$\gamma = \sigma_{12}(\eta N_2 - N_1)\Gamma_s, \tag{24}$$

where $\gamma$ is the small-signal gain coefficient and $L$ is the doped fiber length.

From now on if we want to learn how to change the performance of the gain $G(\lambda)$ in the configuration of 980 nm pumping, we are going to take Eq.(18) and Eq.(19) for $N_1$ and $N_2$ given in Eq.(24), respectively. In an opposite manner, pumping at 1470 nm, we are going to use Eq.(6) and Eq.(7). It is well known that the optimum erbium-doped fiber amplifier length for a maximum gain increases with pump power and decreases with the signal power [12]. For a given signal power and various pumping powers at the wavelengths of 980 *nm* and 1470 *nm*, optimum gains corresponding to optimum lengths from Ref. [13-15] are calculated in the temperature range from 0 to +50 ºC. In addition, the obtained results from this theoretical model are compared with the experimental ones in the literature [16-18]. Furthermore, the term of $R_p$ in the expressions for $N_1$ and $N_2$ is given by

$$R_p = \frac{\Gamma_p \sigma_{13} P_p}{h\nu_p A_p}. \tag{25}$$

## 3. Numerical results

Gain characteristics of EDFAs, depending on the temperature and cross-sections, are investigated for a temperature range from 0 to $+50\,^{o}C$. For this aim $S_{12}$, which is required for the numerical calculations, is calculated firstly. As calculating the maximum gain, optimum fiber amplifier lengths are taken as $L = 5,6\ m$ for 1470 *nm* pumping and $L = 4,2\ m$ for 980 *nm* pumping [19]. The obtained results for the signal power of 50 $\mu W$, taking into account the spectral-backward transmitted ASE powers at longitudinal position z, are $S_{12}$ = 2.06 at 1544 *nm* signal wavelength and $S_{12}$ = 4.05 at 1531 *nm* signal wavelength for the pumping configuration of 1470 nm, $S_{12}$ = 2.32 at 1544 *nm* signal wavelength and $S_{12}$ = 5.34 at 1531 *nm* signal wavelength for the other configuration. The spectral distribution with respect to the corresponding wavelengths of ASE powers is simulated by means of OptiAmplifier version 4.0 (Figs. 2(a, b, c, d)). The variation of the relative population difference with the normalized pumping rate $R_p \tau$ is illustrated in Fig.3 for 1470 *nm* pumping and in Fig. 4 for 980 *nm*, employing the different temperatures within the temperature range given above. The values of $\lambda_0$ are calculated from [9] as $\lambda_0 = 1523\ nm$ for *1531 nm* signal wavelength and $\lambda_0 = 1525\ nm$ for *1544 nm* signal wavelength.

**Fig. 2 (a, b, c, d).**

**Fig. 3.**

**Fig. 4.**

The numerical values required in this work are summarized in Table 1 and the relevant fiber amplifier parameters used in our calculations are as follows:

$\Gamma_s = 0,8$; $\Gamma_{p,980} = 0,85$; $\Gamma_{p,1470} = 0,72$;

$A_s = 52,6 \ \mu m^2$; $A_{p,980} = 33,6 \ \mu m^2$; $A_{p,1470} = 49,7 \ \mu m^2$;

$\sigma_{13} = \begin{cases} 2,58x10^{-25} \ m^2, & \text{for 980 nm pumping} \\ 1,64x10^{-25} \ m^2, & \text{for 1470 nm pumping} \end{cases}$

$\sigma_{12} = \begin{cases} 5,94x10^{-25} \ m^2, & \text{for 1531 nm signal wavelength} \\ 3,26x10^{-25} \ m^2, & \text{for 1544 nm signal wavelength} \end{cases}$

$\sigma_{21} = \begin{cases} 7,04x10^{-25} \ m^2, & \text{for 1531 nm signal wavelength} \\ 4,80x10^{-25} \ m^2, & \text{for 1544 nm signal wavelength} \end{cases}$

$N = 1,0 x 10^{25} \ m^{-3}$, $\tau = 10 \ ms$.

**Table 1**

For 1544 *nm* and 1531 *nm* signal wavelengths, the signal power is taken as *50 μW*. The fiber lengths *L* are kept short to avoid saturation effects resulting from amplified spontaneous emission.

According to the results obtained in this study, the pumping at 980 *nm* (Fig. 4) shows higher population inversion difference than that of 1470 *nm* (Fig. 3) for all normalized pump rates. For 1470 *nm* pumping configuration, when the temperature is increased from 0 $^o$C to + 50 $^o$C, the population inversion obtained from Eq.(5), which depends on $\beta$ *(T)* and $\eta$ *(T)*, decreases and its value for 1544 *nm* signal wavelength is lower than that for 1531 *nm*. For the pumping configuration of 980 *nm*, if the temperature is changed from 0 $^o$C to + 50 $^o$C, the population inversion from Eq.(17) does not change for both 1544 *nm* and 1531 *nm* signal wavelength although it depends on $\eta$ *(T)*, as shown in Fig. 4.

For the signal power of 50 *μW* and the both signal wavelengths, the obtained gains for the pumping at 1470 *nm* (Figs. 5 (a, b)) are higher than that of 980 *nm* (Figs. 6 (a, b)) in the high-pump-power region ($\geq$ 25 mW). On the other hands, the reverse situation occurs in low-pump-power region (~ 5 – 20 *mW*; Figs. 6 (a, b)). The difference in the gains is due to the greatness of absorption cross section at 980 *nm* pumping than that of 1470 *nm* and the smallness of LP$_{01}$-mode size at 980 *nm* pumping than that of 1470 *nm*.

**Fig. 5 (a, b).**

**Fig. 6 (a, b).**

**4. Conclusions**

The variation in the relative population inversion and the net signal gain with temperature is analytically investigated in the temperature range from 0 to +50 $^o$C. It is seen that the gain in this temperature range increases for the certain pump and the signal wavelengths as the temperature decreases. The temperature dependency of the net signal gain as a function of launched pump power are investigated at the signal wavelengths of 1531 and

1544 *nm*, which are the two peak wavelengths in the signal gain spectrum of $Er^{3+}$-doped aluminosilicate type optical fiber. All figures belonging to the relative population inversion and the net signal gain show that the performance of EDFAs depends fairly on the temperature for 1470 *nm* pumping configuration. However, for 980 *nm,* the gain depends on the temperature yet although the dependence of the relative population inversion upon the temperature is negligible. Consequently, regarding to the results from the rate equation model modified by including temperature and cross sections, we found that the performance of EDFAs is more efficient at the pumping wavelength of 1470 *nm* than that of 980 *nm* for a wide range of temperature and high pump powers.

**Acknowledgement**

This study is supported by Scientific Research Projects Council (SRPC) of Erciyes University; project number FBT-04 -17.

**Figure Legends**

Fig. 1. Energy level scheme representing the three-level gain system.

Fig. 2. Spectral-backward ASE powers at the longitudinal position z (from OptiAmplifier version 4.0 software); **a)** for the signal wavelength of 1531 nm at the pumping configuration of 980 nm, **b)** for the signal wavelength of 1544 nm at the pumping configuration of 980 nm, **c)** for the signal wavelength of 1531 nm at the pumping configuration of 1470 nm, **d)** for the signal wavelength of 1544 nm at the pumping configuration of 14700 nm.

Fig. 3. Relative population difference versus normalized pumping rate for different $\eta$ and $\beta$ values at 1531 *nm* signal wavelength, corresponding to *1470 nm* pumping configuration for the signal power of 50 $\mu W$.

Fig. 4. Relative population difference versus normalized pumping rate for different $\eta$ and $\beta$ values. All of the curves at 1544 *nm* and 1531 *nm* signal wavelengths for the relevant temperature values, corresponding to the pumping configuration of *980 nm* for the signal power of 50 $\mu W$, overlap.

Fig. 5. Net signal gain versus launched pump power for 1470 *nm* pumping configuration in 50 $\mu W$ signal power regime. **a)** at 1531 *nm*, **b)** at 1544 *nm* signal wavelengths, for a temperature range of 0 to 50 $^oC$.

Fig. 6. Net signal gain versus launched pump power for 980 *nm* pumping configuration in 50 $\mu W$ signal power regime. **a)** at 1531 *nm*, **b)** at 1544 *nm* signal wavelengths, for a temperature range of 0 to 50 $^oC$.

Table 1: Required parameters for 1470 *nm* and for 980 *nm* pumping configurations.

**Figures:**

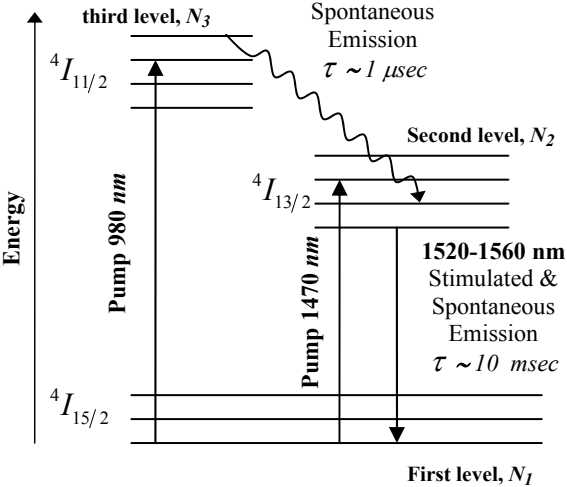

**Fig. 1**

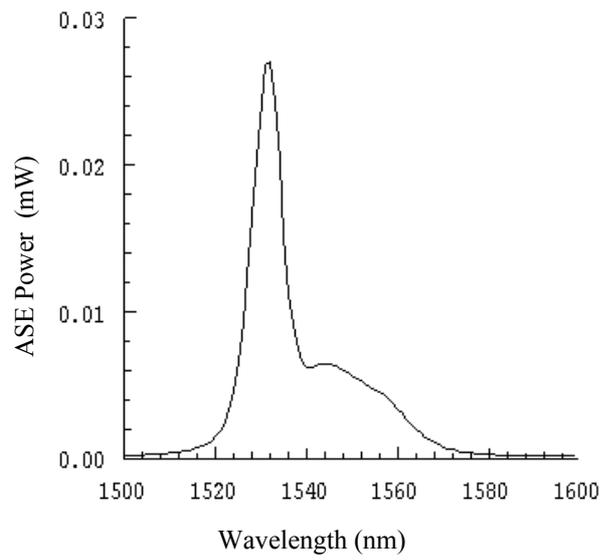

a)

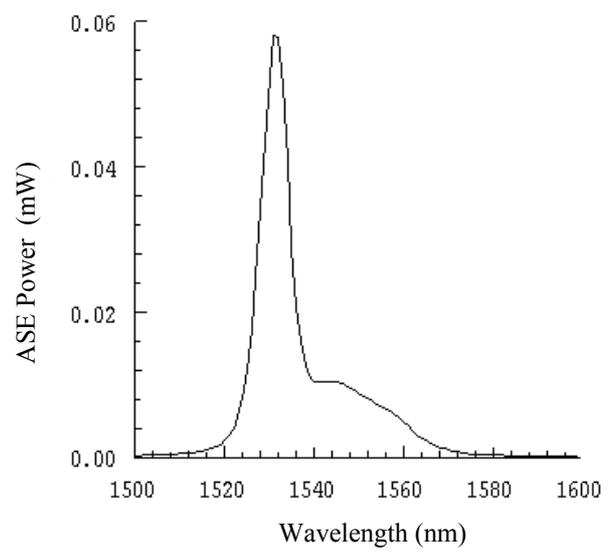

b)

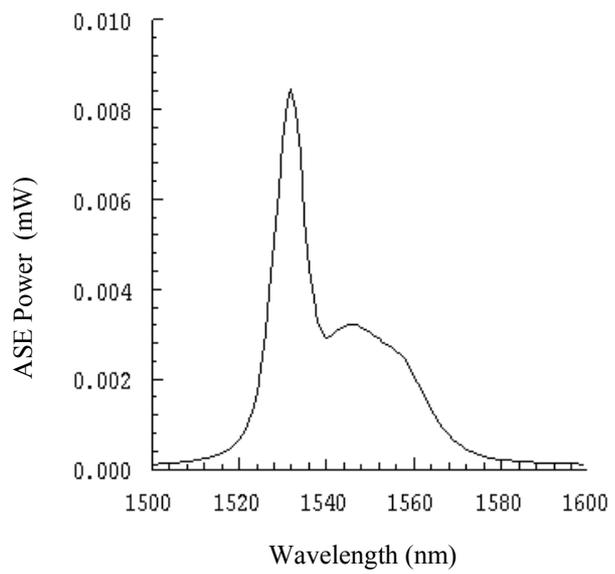

c)

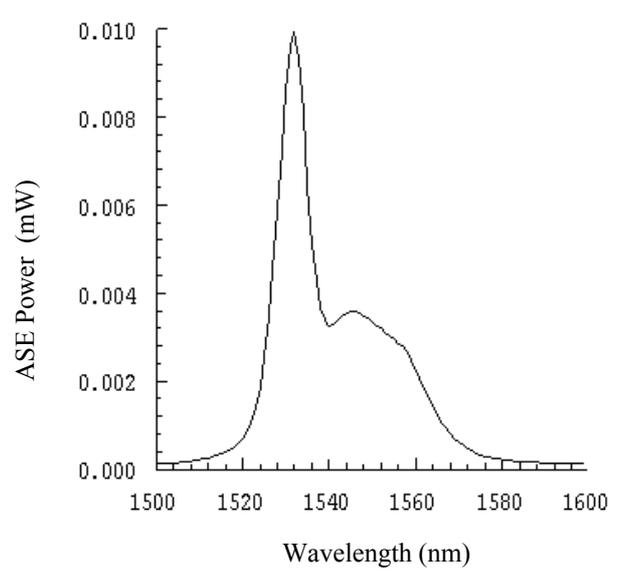

d)

**Fig. 2**

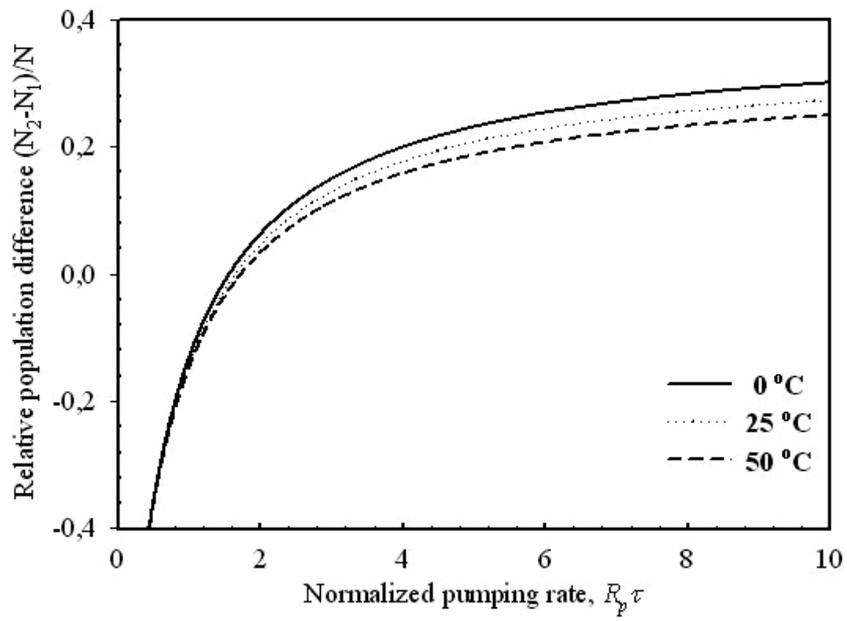

**Fig. 3**

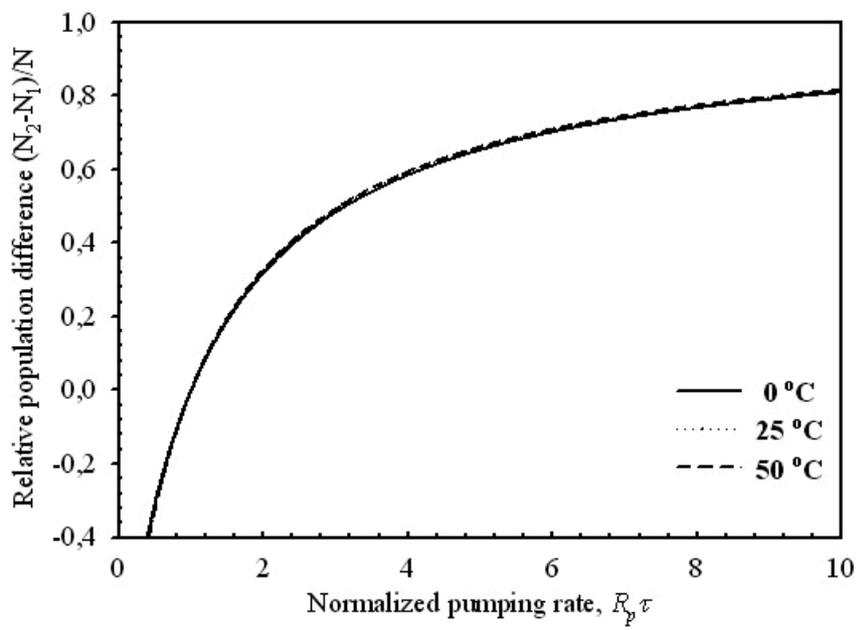

**Fig. 4**

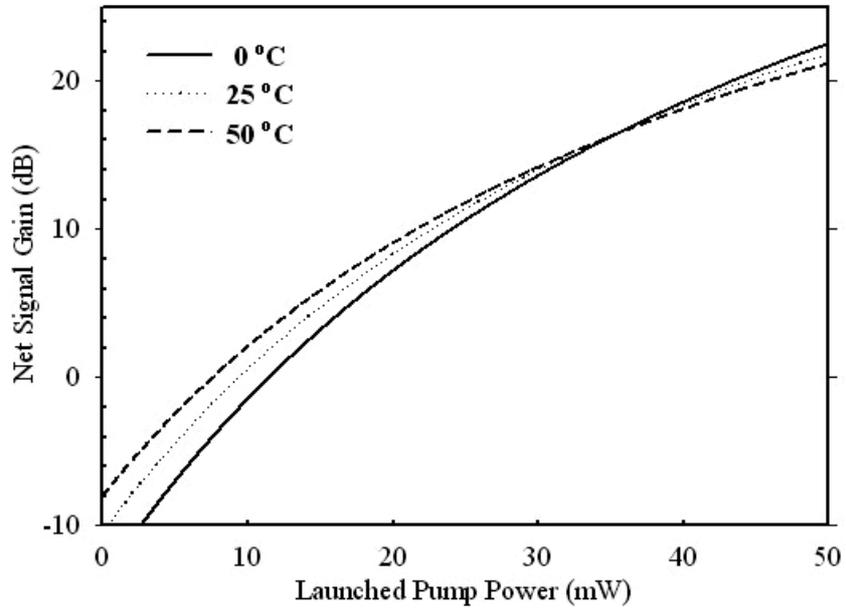

**Fig. 5a**

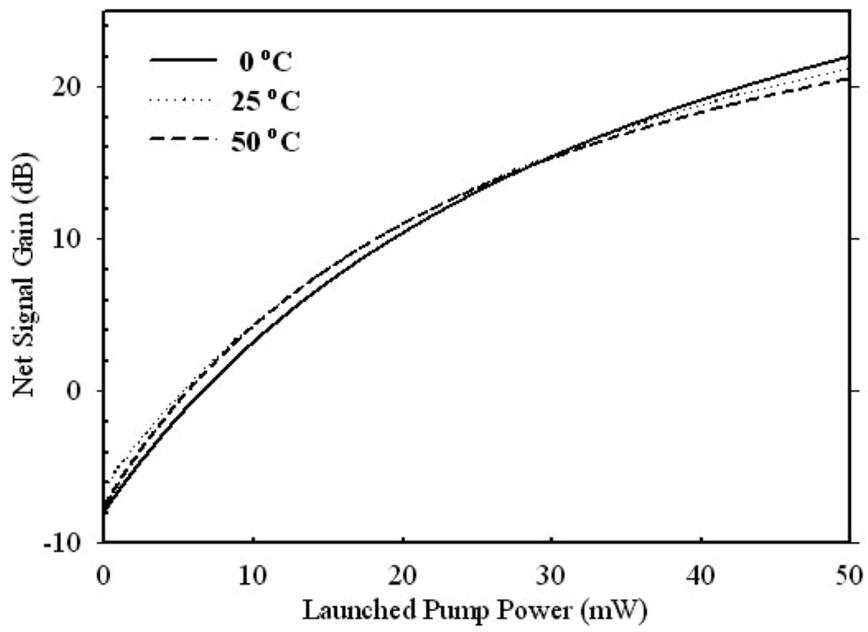

**Fig. 5b**

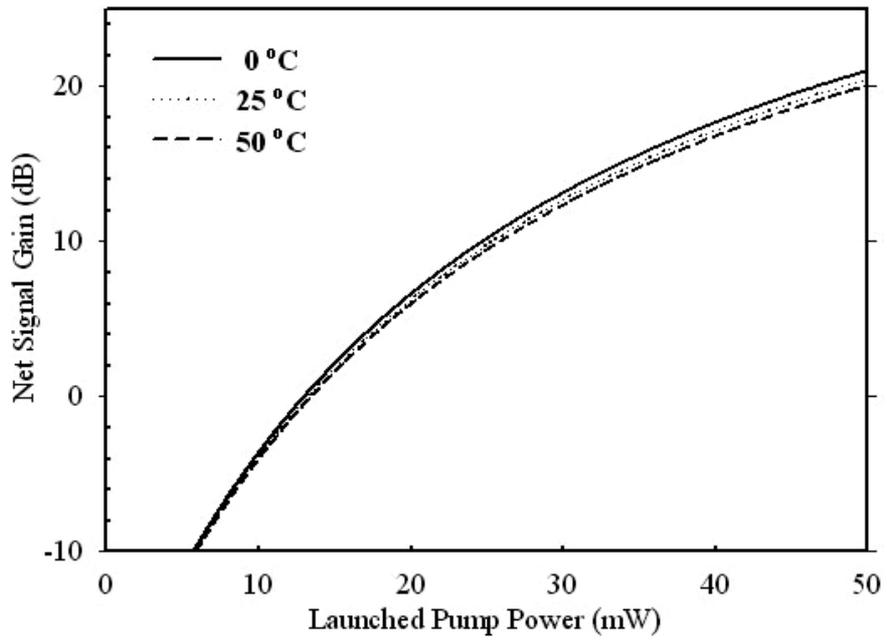

**Fig. 6a**

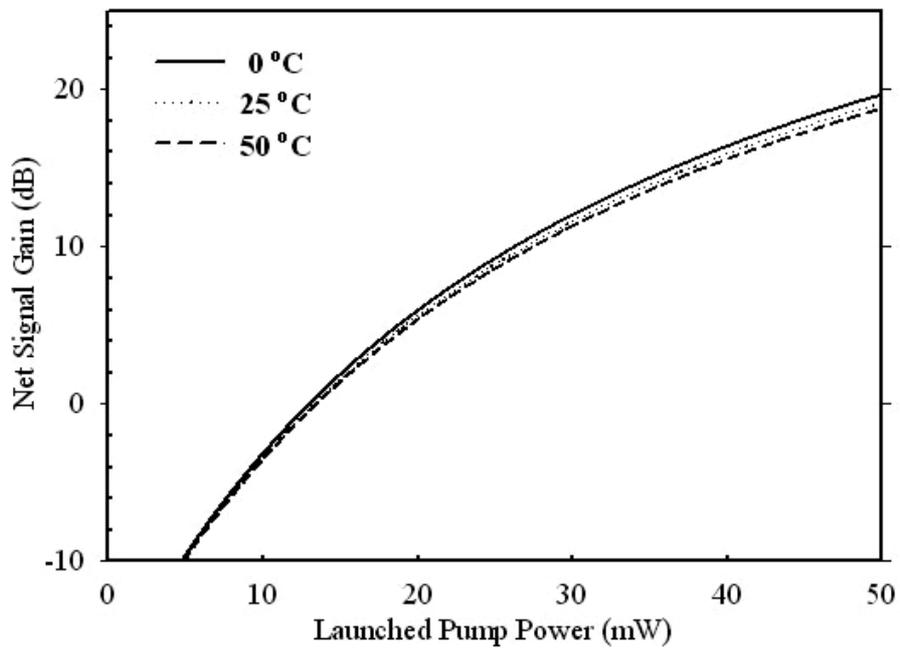

**Fig. 6b**

**Table 1**

| T (°C) | $\beta$ (1470 nm) | $\eta$ (1531 nm) | $\eta$ (1544 nm) |
|---|---|---|---|
| *0* | *0,354* | *1,198* | *1,530* |
| *25* | *0,386* | *1,179* | *1,476* |
| ***50*** | *0,416* | *1,165* | *1,433* |